\documentclass[11pt,a4paper]{article}

\usepackage{jheppub}
  
\usepackage{amsmath}
\usepackage{amsfonts}
\usepackage{amssymb}

\usepackage{graphicx}
\usepackage{float}
\DeclareGraphicsExtensions{.ps}

    \newcommand{\bx}[1][x]{(#1)}
    \newcommand{\btz}[1][x]{(t, #1_3)}
    \newcommand{\bzerox}[1][x]{(0, \mathbf{#1})}

    \newcommand{\wvec}{k}
    \newcommand{\wvecalt}{l}
    \newcommand{\wveci}[1]{\wvec_{#1}}
    \newcommand{\modelabel}{\lambda}

	\newcommand{\Deltax}[1]{\Delta x_{#1}}
	
	\newcommand{\Deltamode}{\Delta\modelabel}

	\newcommand{\LATx}{X}
	\newcommand{\LATy}{Y}
	\newcommand{\LATt}{T}
	\newcommand{\LATxi}[1]{\LATx_{#1}}
	\newcommand{\LATyi}[1]{\LATy_{#1}}
	\newcommand{\LATbx}{(\LATx)}	  	  
	\newcommand{\LATbxp}[1]{(\LATx+#1)}
	\newcommand{\LATbxm}[1]{(\LATx-#1)}
	\newcommand{\LATbzerox}[1][\LATx]{(0, \mathbf{#1})}
	\newcommand{\LATbtz}[1][\LATxi{3}]{(\LATt, #1)}
	\newcommand{\LATbtzp}[2][\LATxi{3}]{(\LATt, #1+#2)}
	\newcommand{\LATbtzm}[2][\LATxi{3}]{(\LATt, #1-#2)}
	\newcommand{\LATbtpz}[1][\LATxi{3}]{(\LATt+0, #1)}
	\newcommand{\LATbtmz}[1][\LATxi{3}]{(\LATt-0, #1)}
	\newcommand{\LATbxdx}[1][\mathbf{\Delta x}]{(\LATx, #1)}
	\newcommand{\LATbdx}{(\mathbf{\Delta x})}	

	\newcommand{\LATwvec}{K}
	\newcommand{\LATwveci}[1]{K_{#1}}
	\newcommand{\LATmodelabel}{\Lambda}
	\newcommand{\LATkas}{\LATwveci{1},\LATwveci{2},\LATmodelabel, s}

	\newcommand{\Id}[1]{\mathbb{I}_{#1}}
	\newcommand{\opgeneric}{\hat{O}}
	\DeclareMathOperator{\erf}{erf}

	\newcommand{\qaverage}[1][\cdot]{\langle#1\rangle}

	\newcommand{\bdown}[1][\mathbf{\wvec}, s]{\hat{b}_{#1}}
	\newcommand{\bup}[1][\mathbf{\wvec}, s]{\hat{b}^{\dagger}_{#1}}
	\newcommand{\ddown}[1][\mathbf{\wvec}, s]{\hat{d}_{#1}}
	\newcommand{\dup}[1][\mathbf{\wvec}, s]{\hat{d}^{\dagger}_{#1}}
	\newcommand{\bdownmode}[1][\wvec_1,\wvec_2, \modelabel, s]{\hat{b}_{(#1)}}

	\newcommand{\dupmode}[1][\wvec_1,\wvec_2, \modelabel, s]{\hat{d}^{\dagger}_{(#1)}}

	\newcommand{\Bferm}{B}
	\newcommand{\BfermN}[1]{\Bferm_{#1}}

	\newcommand{\hmod}{\rho}
	\newcommand{\hphase}{\theta}
	\newcommand{\ferm}{\Psi}
	\newcommand{\Ftens}{F}
	\newcommand{\gauge}{A}
	\newcommand{\Amu}[1][\mu]{\gauge_{#1}}
	\newcommand{\Amuup}[1][\mu]{\gauge^{#1}}
	\newcommand{\elec}[1]{E_{#1}}
	\newcommand{\magn}{B}
	\newcommand{\magni}[1]{\magn_{#1}}
	\newcommand{\jmu}[1][\mu]{j^{#1}}

	\newcommand{\qferm}{\hat{\ferm}} 
	\newcommand{\qfermbar}{\hat{\bar{\ferm}}} 
	\newcommand{\fmodefull}[2][\mathbf{\wvec}, s]{\psi_{#1}^{\left(#2\right)}}

	\newcommand{\fmodeplusneum}{\tilde{\psi}_{+}}
	\newcommand{\fmodeminusneum}{\tilde{\psi}_{-}}
	\newcommand{\fmodez}[2][\wvec_1, \wvec_2, \modelabel, s]{\chi^{(#2)}_{(#1)}}
	\newcommand{\fmodebarz}[2][\wvec_1, \wvec_2, \modelabel, s]{\bar{\chi}^{(#2)}_{(#1)}}

	\newcommand{\Umu}[1][\mu]{U_{#1}}
	\newcommand{\Xphase}{\Theta}

	\newcommand{\DeltaV}{\Delta V}

\title{Simulations of ``Tunnelling of the 3rd Kind"}

	\author[b]{Zong-Gang Mou,}
	\author[a]{Paul M. Saffin,}
	\author[a]{Paul Tognarelli,}
	\author[b]{Anders Tranberg}

	\affiliation[a]{School of Physics and Astronomy, University Park, University of Nottingham,\\ Nottingham NG7 2RD, United Kingdom}
	\affiliation[b]{Faculty of Science and Technology, University of Stavanger, 4036 Stavanger, Norway}

	\emailAdd{zonggang.mou@uis.no}
	\emailAdd{paul.saffin@nottingham.ac.uk}
	\emailAdd{paul.tognarelli@nottingham.ac.uk}
	\emailAdd{anders.tranberg@uis.no}
	

\abstract{We consider the phenomenon of ``tunnelling of the 3rd kind" \cite{third}, whereby a magnetic field may traverse a classically impenetrable barrier by pair creation of unimpeded quantum fermions. These propagate through the barrier and generate a magnetic field on the other side. We study this numerically using quantum fermions coupled to a classical Higgs-gauge system, where we set up a magnetic field outside a box shielded by two superconducting barriers. We examine the magnitude of the internal magnetic field, and find agreement with existing perturbative results within a factor of two. }

\begin{document}

\maketitle

\section{Introduction}
\label{sec:Intro}

Classically, magnetic fields are unable to propagate through a variety of barriers, mirrors, walls, and thick superconductors. At the quantum level, however, the possibility arises that particles may be pair-created on one side of the barrier and propagate through to the other side. Once there, a magnetic field may be (re-)created by the particle pair, leading to the phenomenon dubbed ``tunnelling of the 3rd kind" \cite{third}. 

The assumption is that the pair-created particle species does not see the barrier, and that excitations may propagate freely the distance required before in effect annihilating again. 

A careful demonstration of the effect was made in \cite{third}, with the expected magnetic field beyond the barrier computed in a number of relevant limits (see below). Further applications have since been investigated \cite{third2,third3}.

In this work, we investigate the phenomenon non-perturbatively and numerically on a lattice, simulating quantum fermions coupled to electromagnetism (a $U(1)$ gauge field). The gauge field is in turn coupled to a classical scalar field, which may be engineered to have a large, localized field expectation value, thereby inflicting a large mass on the photons. The scalar field then effectively acts as a barrier, where magnetic fields have a finite penetration depth. But since the fermions are not coupled to the scalar field, they are oblivious (at least to leading order) to its existence. 

We will consider a very specific setup, where two parallel 2-dimensional walls split 3 dimensional space into one ``inside" region, the ``box", and two ``outside" regions, one on either side. We may then effectively reduce the simulations to be one-dimensional, invariant under translations in the directions transverse to the walls. Through a number of technical developments, we can engineer a magnetic field initially outside the box, and by varying the strength and thickness of the barriers, investigate whether the magnetic field inside the box is affected by the presence of quantum fermions in the system. 

A secondary goal of this work, is to demonstrate that methods of classical/quantum field theory including fermions in real-time \cite{Aarts1,Aarts2}, may be used to model real-life experiments numerically (see also \cite{Berges3}). Lattice fermions have been implemented in real time before, but the overwhelming approach has been to consider imaginary time, usually with the view of computing statistical expectation values averaged over many random configurations, in an infinite system (periodic boundary conditions, taking the infinite volume limit). Here, we instead consider a finite system, and address many of the technical difficulties associated with boundary conditions, stability and real-time sources, but using the language and formalism of relativistic quantum fields on a cubic lattice.

Quantum real-time fermions are numerically very expensive, since they involve solving for $2N$ distinct mode functions, with $N$ the lattice size. A ``cheaper" version, taking advantage of a statistical averaging procedure rather than the complete set of mode function was introduced in \cite{Borsanyi1} and applied to a number of phenomena \cite{Berges1, Berges2, Saffin1,Saffin2,Saffin3,Saffin4}. For our purposes here, this approach does not reduce the numerical effort sufficiently at the precision required, and we revert to the ``all-modes" implementation of \cite{Aarts1,Aarts2}.

In Section \ref{sec:Model}, we introduce our model and definitions, and the field equations for classical and quantum fields. In Section \ref{sec:Setup}, we specialise to our setup of choice, describing boundary conditions, the implementation of the box and our observables. In Section \ref{sec:Analytic}, we describe the phenomenon of tunnelling of the 3rd kind, and provide some theoretical background based on \cite{third}, adapted to our setup. Section \ref{sec:MagTrans} contains our numerical tests, comparing the classical and classical plus quantum system in search of a tunnelling signal. Section \ref{sec:Conc} contains our conclusions. The details of the lattice implementation may be found in the Appendices \ref{sec:LattImp}-\ref{sec:AppRenorm}.

%
%
%

\section{Model}
\label{sec:Model}

We consider a model including a complex scalar $\phi$ (the Higgs field), and $U(1)$ gauge field $A_\mu$ (electromagnetism) and a charged fermion field $\Psi$. Writing the scalar in polar form $\phi= \rho \exp(i\theta)$ and keeping only the phase as dynamical, we have the action (signature $(-+++)$)
\begin{multline}
S = \int c\,dt\,d^3\mathbf{x} \bigg[-\frac{1}{4\mu_0}\Ftens_{\mu\nu}\bx\Ftens^{\mu\nu}\bx - \frac{1}{2}\hmod^2\bx D_{\mu}\hphase\bx D^{\mu}\hphase\bx- j_{\mu}\bx\Amuup\bx \\
 -\bar{\ferm}\bx\gamma^{\mu}D_{\mu}\ferm\bx - \frac{mc}{\hbar}\bar{\ferm}\bx\ferm\bx \bigg] \mbox{.}
\end{multline}
As usual, $\bar{\ferm}\equiv i\ferm^{\dagger}\gamma^0$ and we have included an external source $j_\mu(x)$. At this stage, we keep the fundamental constants $c$, $\hbar$, $\mu_0$ explicit. The covariant derivatives are then
\begin{equation}
D_{\mu}\hphase = \partial_{\mu}\hphase\bx - \frac{e}{\hbar}\Amu\mbox{, }\quad
D_{\mu}\ferm = \left(\partial_{\mu} - i\frac{q}{\hbar}\Amu\right)\ferm \mbox{,}
\end{equation}
where the gauge-Higgs coupling is determined by the usual electron charge $e$, but where the fermion is allowed a general charge $q$.
We employ the Weyl representation of the fermion algebra throughout.  

Variation of the action and imposing the temporal gauge yields the equations of motion:
\begin{align}
	\label{eq:EOM_gaus_full3d_class}
	\frac{1}{\mu_0c}\Sigma_i\partial_i\elec{i}\bx + \frac{e}{\hbar}\hmod^2\bx D^0\hphase\bx + i\frac{q}{\hbar}\bar{\ferm}\bx\gamma^0\ferm\bx - \jmu[0]\bx &= 0 \mbox{,}\\
	\label{eq:EOM_elec_full3d_class}
	\frac{1}{\mu_0c}\partial_{0}\elec{i}\bx + \frac{1}{\mu_0}\Sigma_j\partial_j\Ftens_{ji}\bx +\frac{e}{\hbar}\hmod^2\bx D^{i}\hphase\bx + i\frac{q}{\hbar}\bar{\ferm}\bx\gamma^{i}\ferm\bx - \jmu[i]\bx &= 0 \mbox{,}\\
	%
	%
	\label{eq:EOM_ferm_full3d_class}
	\left(\gamma^{\mu}D_{\mu} + \frac{mc}{\hbar}\right)\ferm\bx &= 0 \mbox{,}\\
	\label{eq:EOM_phas_full3d_class}
	\partial_{\mu}\left(\hmod^2\bx D^\mu\hphase\bx\right) &= 0\mbox{,}
\end{align}
with $\elec{i}\bx = -c\partial_0\Amu[i]\bx$.  The first line is Gauss' Law which is satisfied at all times, if enforced initially. The subsequent three equations determine the time evolution of the gauge, scalar and fermion fields. 


At this stage, promoting $\ferm$ to a quantum-field operator and forming the expectation value of the resultant operator bilinears $i\frac{q}{\hbar}\bar{\ferm}\bx\gamma^0\ferm\bx$ and $i\frac{q}{\hbar}\bar{\ferm}\bx\gamma^{i}\ferm\bx$, yields a set of semi-quantized equations of motion in a way similar to \cite{Aarts1, Aarts2}. By this we mean that the linear fermion equation is to be solved as an operator equation (see below), and that the classical gauge and scalar equations are solved non-linearly with the resulting fermion bilinears inserted, computed with the fermion solution. 

We implement these evolution equations on a finite volume, discretized lattice, where we deal with the fermion doubling phenomenon by introducing a Wilson term (see for instance \cite{Robust}). 

\subsection{Quantum fermions}
\label{sec:QuantFerm}

We expand the fermion operator field on Gaussian modes\footnote{Note that since the fermion equation is linear, fermions are always Gaussian, also in an interacting theory, and can be expanded in mode functions.} $\fmodefull{U}$ and $\fmodefull{V}$: 
\begin{equation}
\label{eq:mode-expansion}
\qferm\bx = \sqrt{\hbar c}\,
\sum_s \int \frac{\mathrm{d}^3\mathbf{\wvec}}{(2\pi)^3}\left(\bdown\fmodefull{U}\bx + \dup\fmodefull{V}\bx\right),
\end{equation}
During the time-evolution, each mode independently satisfies the fermion equation of motion (\ref{eq:EOM_ferm_full3d_class}).  For each value of momentum ${\bf k}$, the spin variable $s$ takes two values for each of the two modes $U$ and $V$. 

The $\{\bup\}$ $\{\bdown\}$ are the creation and annihilation operators respectively for the fermionic particle and $\{\dup\}$ $\{\ddown\}$ are for the anti-particle. There is one for each mode function (value of ${\bf k}$, $s$, $U/V$), they are time-independent and satisfy:
\begin{align}
\label{eq:commutator_ladder_wavevector}
\{\bdown, \ddown[\mathbf{\wvecalt}, r]\} 
=
\{\bup, \ddown[\mathbf{\wvecalt}, r]\} = 0
\qquad
\{\bdown, \bup[\mathbf{\wvecalt}, r]\} =  
\{\ddown, \dup[\mathbf{\wvecalt}, r]\} = 
\left(2\pi\right)^3\delta^3(\mathbf{\wvec} - \mathbf{\wvecalt})\delta_{sr}\mbox{.} 
\end{align}
This defines the initial state of the fermions to be the free-field vacuum. The (lattice) implementation of the quantum fermions closely follow \cite{Aarts1}, and details will given in the Appendix \ref{sec:AppFerm}.

\section{Setup of the system}
\label{sec:Setup}

\begin{figure}[H]
  \centering
    \includegraphics[width = 0.65\textwidth]{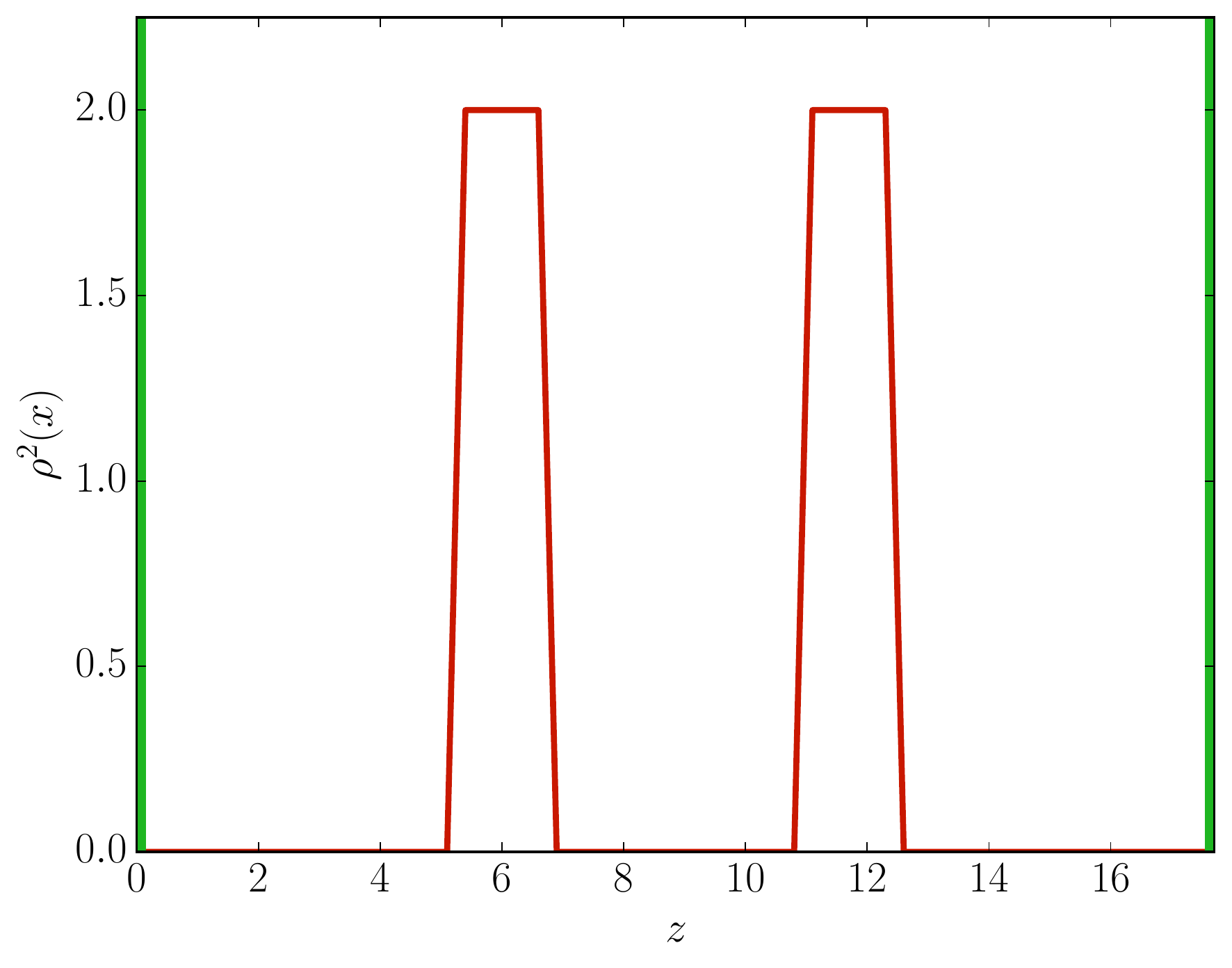}
\caption{The setup of the system with barriers (red) and boundaries (green). Here, $\eta=1$.}
  \label{fig:barriers}
\end{figure}

We consider a 3-dimensional space, where along the $z$-axis two parallel 2-dimensional superconductors are placed, extending in the $x$-$y$ direction. The $z$-direction is taken to be the finite interval $[0,L]$, and the two superconductors are placed symmetrically around $z=L/2$. We choose explicitly 
\begin{align}
\hmod^2\bx = \hmod^2(z) \equiv \eta^2 \Bigg[2 &- \tanh\left(\frac{z + S/2 + d/2 -L/2}{\delta}\right)\tanh\left(\frac{z + S/2 - d/2 - L/2}{\delta}\right) \nonumber\\
&-\tanh\left(\frac{z - S/2 + d/2 - L/2}{\delta}\right)\tanh\left(\frac{z - S/2 - d/2 -L/2}{\delta}\right)\Bigg]\mbox{.}
\label{eq:higgsprofile}
\end{align}
This amounts to two barriers of amplitude $\rho^2=2\eta^2$ at positions $L/2 \pm S/2$, of width $d$ with $\delta$ parametrising how abruptly the scalar field drops from its maximum value to zero, Fig. \ref{fig:barriers}. In essence, we have created a box of width $S$ extending indefinitely in the $x$ and $y$ directions.
This implies that the system may essentially reduce to a one-dimensional problem in the $z$-direction.  We impose periodic boundary conditions in $x$ and $y$ and Neumann boundary conditions in $z$. 

The current $j_\mu$ is introduced to create and sustain an electromagnetic field on the outside of the box. We will choose it to only have a component in the $x$-direction
\begin{equation}
j_\mu(x) = j_1(t,{\bf x}).
\end{equation}
This enforces that $A_2=A_3=0$, if we in addition take $\theta=0$ initially, which is allowed by gauge invariance. Setting the external current aligned only in the $x$-direction further determines the orientation of the electric-field purely along this direction and a magnetic field aligned in the $y$-direction.  All these considerations together therefore impose 
\begin{equation}
\Amu[1]\bx = \Amu[1](t,z)\mbox{,}\quad
E_1(x)= E_1(z,t),\quad 
B_2(x)= B_2(z,t),\quad
\jmu[1]\bx = \jmu[1](t,z)\mbox{,}
\end{equation}
while all further components vanish.  With these simplifications, we retain a single dynamical equation to be solved for $A_1(z,t)$,
\begin{equation}
\label{eq:EOM_elec_ansatz_quantum}
\frac{1}{\mu_0c}\partial_0\elec{1}(t,z) + \frac{1}{\mu_0}\partial^2_3\Amu[1](t,z) -\left(\frac{e}{\hbar}\right)^2\hmod^2(z)\Amu[1](t,z) + i\frac{q}{\hbar}\qaverage[T\,\qfermbar(x)\gamma^1\qferm(x)] - j_1(t,z) = 0,
\end{equation}
and a set of auxiliary equations, that are trivially satisfied with $\theta=0$, and need not be solved explicitly,
\begin{align}
\label{eq:EOM_gauss_ansatz_theta_quantum}
-\frac{e}{\hbar}\hmod^2(z)\partial_0\hphase(t,z)+i\frac{q}{\hbar}\qaverage[T\,\qfermbar\bx\gamma^0\qferm\bx]&=0,\\
\label{eq:EOM_curr_ansatz_theta_quantum}
\frac{e}{\hbar}\hmod^2(z)\partial_{2,3}\hphase(t,z)+i\frac{q}{\hbar}\qaverage[T\,\qfermbar(x)\gamma^{2,3}\qferm(x)]&=0,\\
\left(\hmod^2(z)\left[\partial^2_0 - \partial^2_3\right] - 2\hmod(z)\partial_3\hmod(z)\partial_3 \right)\hphase(t,z) &= 0\mbox{.}
\end{align}
The quantum fermion correlator in (\ref{eq:EOM_elec_ansatz_quantum}) is determined through the operator expansion in a mode-ansatz described in section \ref{sec:Symmetry}. We have made explicit the time-ordering in the fermion bilinear expectation values. This will however not play a role, since the fermion fields are Gaussian\footnote{In a full quantum theory, further non-local diagrams could appear in a Schwinger-Dyson expansion including quantum scalar/gauge internal lines. These appear in our treatment through solving the fermion mode functions in the non-trivial time dependent scalar/gauge background.}.

We will refer to the fermion bilinear appearing in the $A_1$ equation (\ref{eq:EOM_elec_ansatz_quantum}) as the ``fermion current", and to the combination $(e/\hbar)^2 \rho^2(z) A_1(z)$ as the ``Higgs current".

\subsection{External current}
\label{sec:extcurrent}

In order to generate a stationary external magnetic field, we introduce a Gaussian, time-dependent current on both boundaries $z=0,L$, explicitly,
\begin{equation}
j_1\bx =J_{\rm max} \Theta(t)\left(\exp\left(-\frac{z^2}{2\sigma^2}\right) -\exp\left(-\frac{(z+L)^2}{2\sigma^2}\right)\right),
\end{equation}
in terms of a rise function
\begin{equation}
\label{eq:LAT_current_time_variation}
\Theta(t) = \frac{1}{2}\tanh\left(\vartheta\right) + \frac{1}{2} \mbox{, \quad} \vartheta = \tan\left(\frac{\pi t}{\tau} - \frac{\pi}{2}\right),
\end{equation}
with $m\tau=80$, $\sigma=0.8/\sqrt{2}$ and $J_{\rm  max}=0.5$.  Care must be taken on the lattice to locate the peaks midway between the lattice-sites.  The explicit expression is provided in Appendix \ref{sec:AppEOM}, this choice proving important for adequate convergence.  

Because we want a stationary magnetic field, we found that in addition to having the current on, we needed to introduce a little damping to the gauge field dynamics to extract energy. We therefore added the term
\begin{equation}
\frac{\zeta}{\mu_0c}\partial_0\elec{1}\bx\mbox{, } \quad \zeta=\frac{6\pi}{35}
\end{equation}
to Eq.~(\ref{eq:EOM_elec_ansatz_quantum}).

\subsection{Simple estimates}
\label{sec:Analytic}

\begin{figure}[H]
\centering
    \includegraphics[width = 0.65\textwidth]{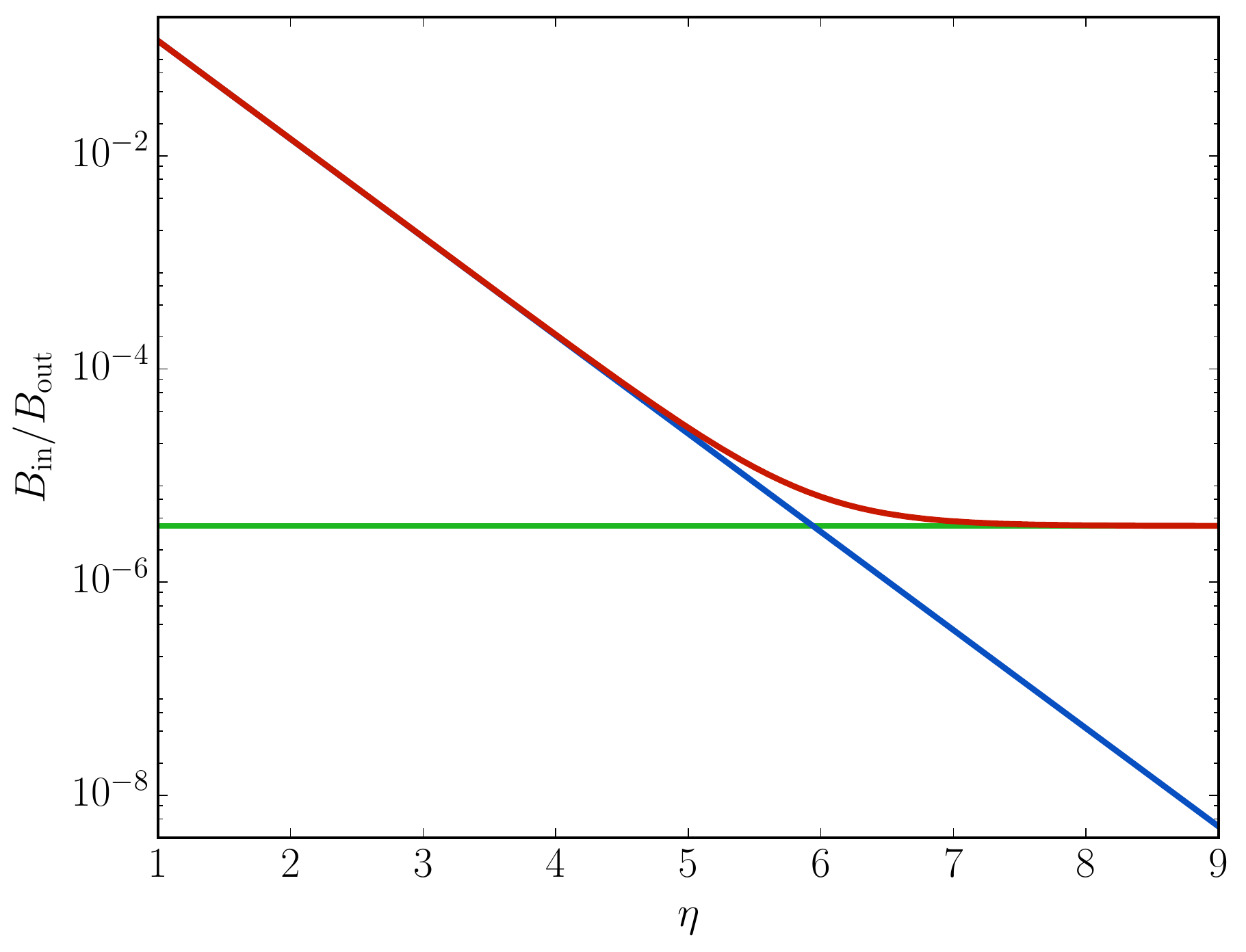}

\caption{Comparing the classical (blue) transmission to the expected fermion signal (green) and their sum (red), using $q=0.3$ and $md=1.5$, as a function of $\eta$.}
\label{fig:Gies}
\end{figure}

A similar physical system, except with one wall, was considered in \cite{third}, where a constant external magnetic field $B_{\rm out}$ was seen to induce a constant magnetic field $B_{\rm in}$ inside the box, of magnitude ($\hbar=c=1$)
\begin{equation}
\frac{|B_{\rm in}|}{|B_{\rm out}|}= \frac{q^2}{24\pi^2} g(md),
\end{equation}
where $m$ and $q$ are the mass and charge of the fermions, $d$ is the thickness of the wall and $g(md)$ is a function to be evaluated numerically
\begin{equation}
g(md)= \frac{1}{2}\int_1^\infty \frac{d\tau}{\tau^4}\sqrt{\tau^2-1}\left(1+2\tau^2\right)\exp[-2 (md) \tau].
\end{equation}
Since we have tunnelling from both sides and the walls are only a small distance from each other, we will allow for an uncertainty factor between one and two to this expression when comparing to our simulations.

In order to discern an effect of including fermion transmission through the wall, we need to compare this to the classical transmission of the gauge field. From the equation of motion, we see that for a stationary field without fermions, and with no external source inside the wall and box, we have
\begin{equation}
\partial_3^2 A_1 = \mu_0\left(\frac{e}{\hbar}\right)^2\rho^2A_1.
\end{equation}
Solving this for a step function of height $2\eta^2$ (a limiting case of the profile defined in (\ref{eq:higgsprofile})), we find trivially ($e=\hbar=1$)
\begin{equation}
A_{1,\rm wall}(t,z) = A_{1, \rm out}e^{-\sqrt{2}\eta z},\quad |B_2(t,z)| = |\partial_3A_{1,\rm wall}(t,z)|= |B_{2,\rm out}|e^{-\sqrt{2}\eta z},
\label{eq:exp_est}
\end{equation}
where $A_{1,\rm out}$ is the gauge field at the outer edge of the wall, and we have imposed continuity of $A_1$ and $B_2$ at the edge of the wall.
We will see that the strict exponential dependence on $\eta$ is not reproduced numerically. We do not have a strict step function, there is a continuous current at the boundary and a damping term in the bulk, so that although the field value is stationary, the shape of the numerical solution may not be exactly as described here. 

In Fig. \ref{fig:Gies}, we show the predicted fermion and classical components  of transmission and their sum as a function of the barrier height $\eta$, given a wall thickness of $md=1.5$. We see that we should expect the fermion contribution to appear at $\eta>6$. But for weaker barriers, we expect it to be swamped by the classical contribution.

\subsection{Symmetry and mode expansion}
\label{sec:Symmetry}

Symmetry considerations further imply a separable ansatz for the fermion mode function, where the $x$, $y$, $z$-dependence enters as the product of three mode functions. In $x$ and $y$ these are simply the free-field, periodic plane-wave solutions. In $z$ they are some unknown functions to be determined numerically. We write
\begin{align}
\fmodefull{U}\bzerox &= e^{i\wvec_1x}e^{i\wvec_2y}\fmodez{U}(0,z), \\
\fmodefull{V}\bzerox &= e^{-i\wvec_1x}e^{-i\wvec_2y}\fmodez[-\wvec_1, -\wvec_2, -\modelabel, s]{V}(0,z), 
\end{align}
where the plane wave solutions in $x$ and $y$ are labelled by the momenta $k_1$ and $k_2$, and the solution in the $z$-direction is labelled by a generic number $\lambda$, which in the plane-wave case would be a momentum $k_3$. In general $\lambda$ just labels a complete set of solutions to the actual equation of motion in $z$. $s$ denotes sum over spin, and we have defined both $U$ and $V$ mode functions. 

The field operator may accordingly be expanded in these functions, with as coefficients creation/annihilation operators $b$ and $d$ ($b^\dagger$, $d^\dagger$): 
\begin{multline}
\qferm\bx = \sqrt{\hbar c} 
\,\sum_s \int \frac{d\wvec_1}{(2\pi)}\frac{d\wvec_2}{(2\pi)}\frac{d\modelabel}{(2\pi)}\Big[\bdownmode e^{i\wvec_1x}e^{i\wvec_2y}\fmodez{U}(t,z) \\
+ \dupmode[-\wvec_1, -\wvec_2, -\modelabel] e^{i\wvec_1x}e^{i\wvec_2x}\fmodez{V}(t,z)\Big]\mbox{.}
\end{multline}


\subsection{Initial Conditions}
\label{sec:InitCond}

We initialize the system in the classical vacuum background
\begin{align}
\elec{1}(0,z) =  0, \quad 
\Amu[1](0,z) =  0,
\end{align}
with the fermion vacuum quantum state. 


This allows us to also introduce plane-wave solutions in $z$ initially, with the label $\lambda\rightarrow k_3$, subject to the Neumann boundary conditions in Appendix \ref{sec:AppBound}. 
 The combination
\begin{equation}
\label{eq:LAT_init_fermU}
\fmodeplusneum\bx = e^{i(\wveci{1}x + \wveci{2}y)}\left(e^{i\wveci{3}z}U_{\mathbf{\wvec},s} +  i\gamma^5\gamma^3e^{-i\wveci{3}z}U_{\mathbf{\wvec},s}\right),
\end{equation}
provides the initial solution associated to the positive-energy particles; likewise, the linear combination 
\begin{equation}
\label{eq:LAT_init_fermV}
\fmodeminusneum\bx = e^{-i(\wveci{1}x + \wveci{2}y)}\left(e^{-i\wveci{3}z}V_{\mathbf{\wvec},s} +  i\gamma^5\gamma^3e^{i\wveci{3}z}V_{\mathbf{\wvec},s}\right),
\end{equation}
provides the initial solution associated to the negative-energy anti-particles.  Some care must be taken to ensure that the boundary conditions are observed also on a finite-volume discretized lattice. Details of this may be found in Appendix \ref{sec:AppFermBound}.

\subsection{Renormalization}
\label{sec:Renorm}

Fermion bilinears evaluated in the vacuum are in principle divergent, and we need to renormalise the gauge field equation. We do this by performing a wave-function renormalization, essentially by adding a counterterm to the classical electromagnetic field tensor in the action. This results in the addition to the gauge field equation of motion, Eq.~(\ref{eq:EOM_elec_ansatz_quantum}), of
\begin{equation}
\alpha\left(\frac{1}{\mu_0c}\partial_0\elec{1}(t,z) + \frac{1}{\mu_0}\partial^2_3\Amu[1](t,z)\right),
\end{equation}
with $\alpha$ playing the role of counterterm. We can tune this parameter to give convergence of the dynamics as we vary lattice spacing. This is described in some detail in Appendix \ref{sec:AppRenorm}, but for most of the simulations presented here, $\alpha=0$, corresponding to us defining the renormalization point at the scale $dx=0.3$ (see below). We are not taking the continuum limit (requiring a larger numerical lattice out of our reach), and the renormalization procedure was used as a check that we have the lattice spacing effects under control.

\section{Do we observe tunnelling of the 3rd kind?}
\label{sec:MagTrans}

We will now introduce natural units $\hbar = c = 1$.  The 3-dimensional lattice has spacings $dx=dy=dz= 0.3$, in units where the fermion mass $m$ is unity and we use the time-step $dt= 0.002$.  We choose the fermion-gauge coupling to be $q=0.3$ and Higgs-gauge coupling $e$ to be unity. The Wilson coefficient is chosen to be $r_W = 0.5$, and we checked that this gives good control of the fermion doublers. There are $N_x=N_y=32$, $N_z=60$ lattice sites, which is the largest attainable with our numerical resources. This means that the physical length in the $z$-direction is $mN_z dz =18$, split up into $5.4$ left and right of the barrier, $1.5$ in each barrier and $4.2$ inside the box (Fig. \ref{fig:barriers}).

In each simulation, we start in the vacuum and gradually turn on the external source, as described, from zero to the maximum value $J_{\rm  max}=0.5$.  The turning on takes $m\tau=80$, and we allow the system to settle until $mt=150$. This yields a
very accurately uniform magnetic-field external to the superconducting-barrier, on both sides of the box.    

\begin{figure}[H]
  \centering
    \includegraphics[width = 0.48\textwidth]{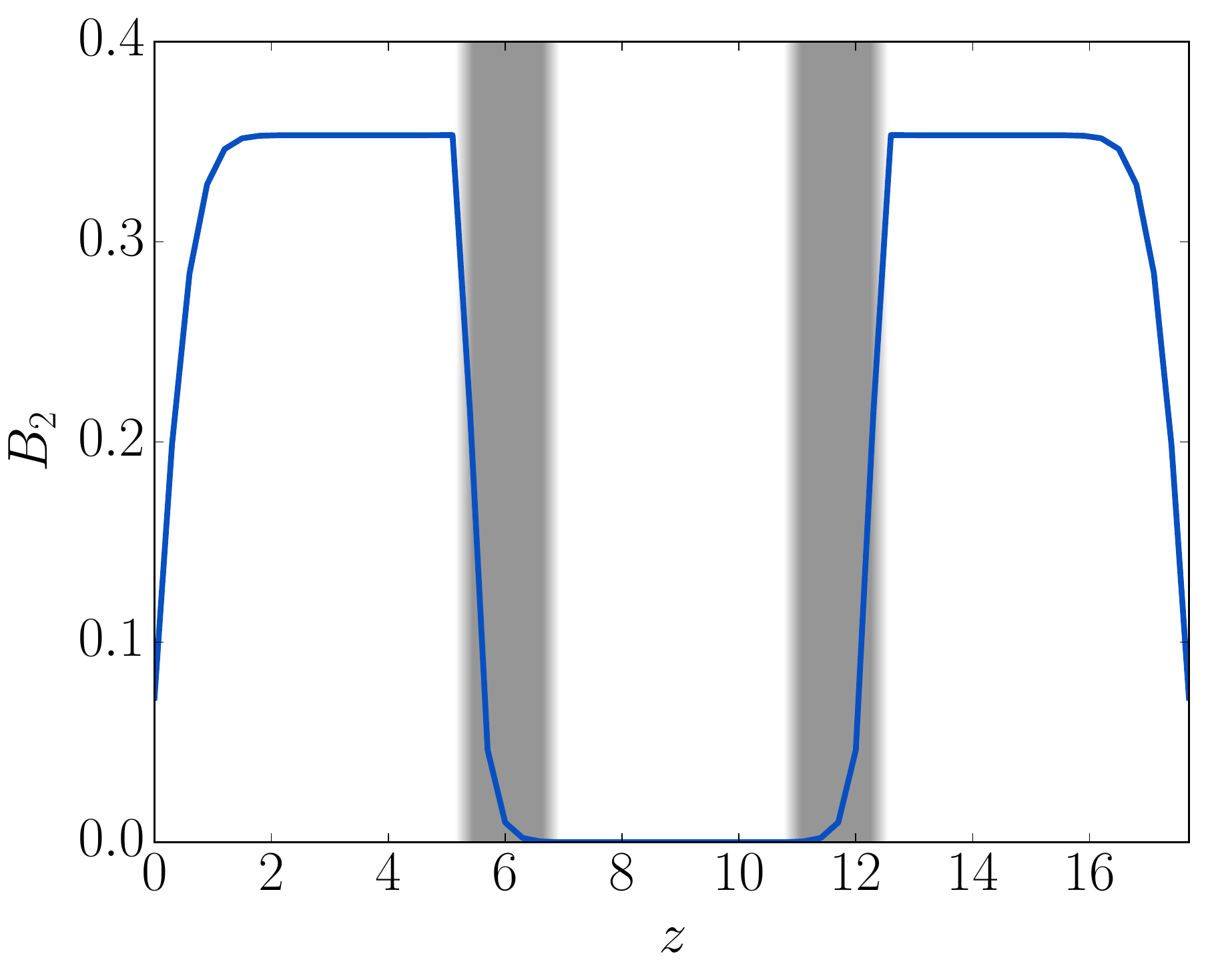}
     \includegraphics[width = 0.48\textwidth]{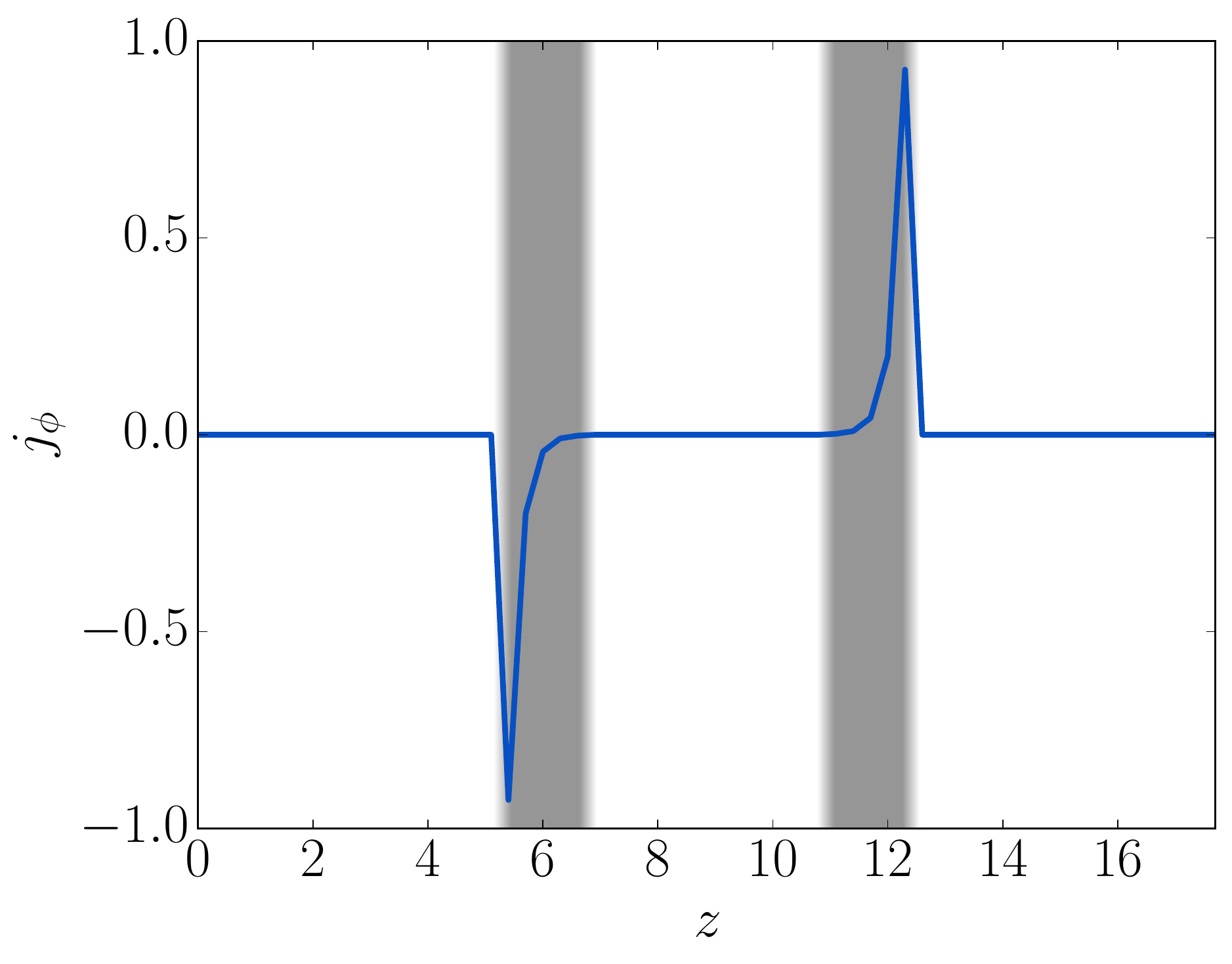}
\caption{The stationary magnetic field in a typical simulation, after the current is turned on (left). The grey lines represent the superconducting barriers. The corresponding Higgs current (right). $q=0.3$, $\eta=4$, $d=1.5$.}
  \label{fig:bosfields}
\end{figure}

In Fig. \ref{fig:bosfields}, we show the magnetic field $B_2(z)$ as the stationary state is achieved (left panel). The superconducting barriers are shown as grey bands, and we see that the magnetic field grows near the boundary where the current is, and as it reaches the wall decays exponentially to leave very little on the inside of the box. We also see that the wall is well separated from the boundary effects of the external current.

\begin{figure}[H]
	\centering
	\includegraphics[width = 0.48\textwidth]{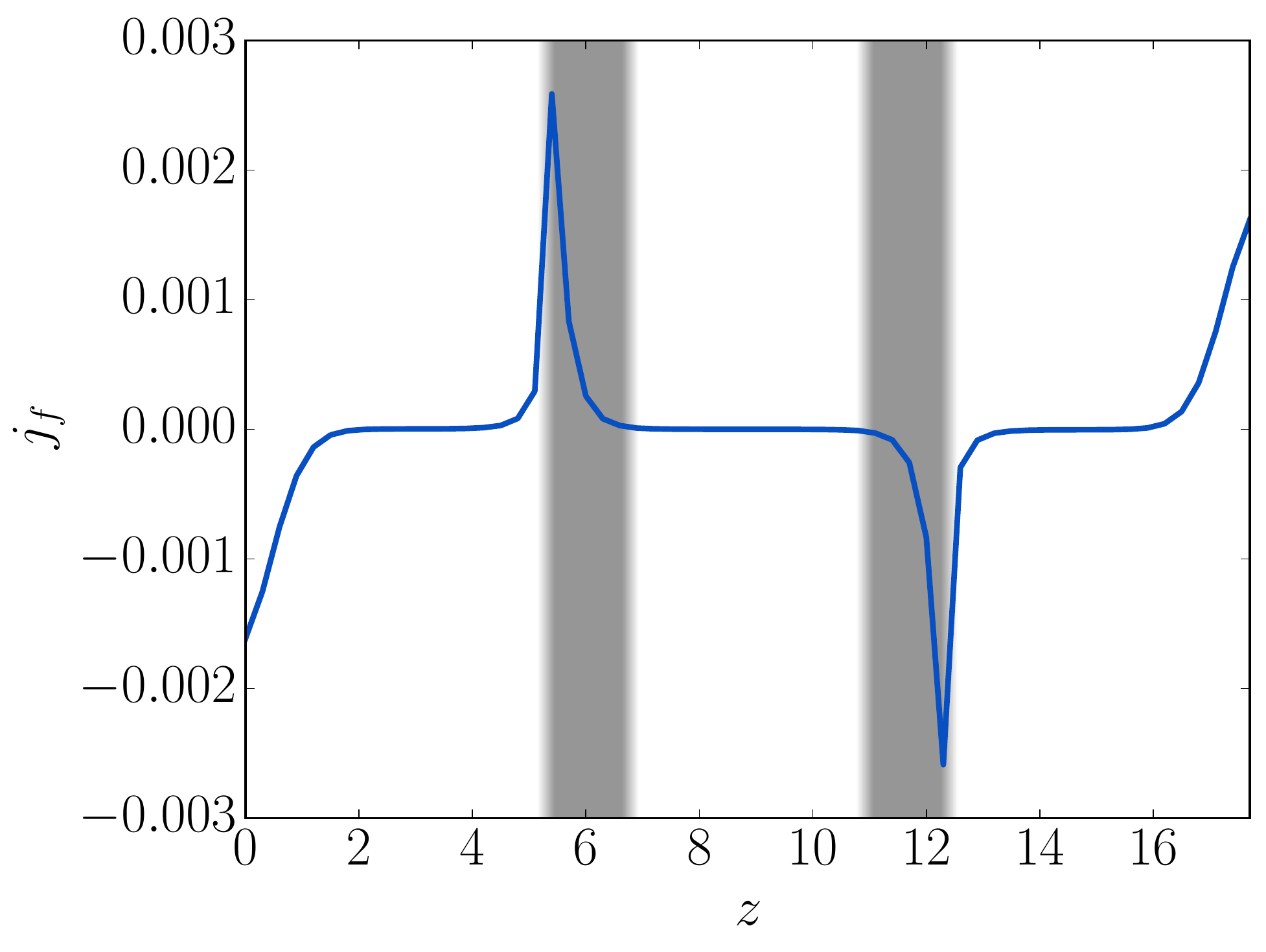}
	\caption{The fermion current in a typical simulation. $q=0.3$, $\eta=4$, $d=1.5$.}
	\label{fig:fermcurr}
\end{figure}

We also show the Higgs current (right panel of Fig. \ref{fig:bosfields}), which has large peaks at the outer edge of the walls, that decay exponentially to near zero inside the box. At the scale displayed here, there is no apparent difference from including or omitting (not shown) the fermions. In this example, the walls have width $d=1.5$, and the amplitude of the walls is $\eta=4$. The parameter $\delta$ is much smaller than the lattice spacing, so the wall is close to a step function. Note that essentially the Higgs current is $\propto A_1\rho^2$, and so the decay is due to the decay of $A_1$, and the peak follows from the near-step function of $\rho$. Fig.~\ref{fig:fermcurr} shows the fermion current in the same simulation, and we again see peaks, also decaying through the walls.

The fermion current at the boundary notably acts in accordance with Lens' Law to oppose the magnetic-flux inducing the current. The current is hence aligned in the opposite direction to the external current.  Increasing the wall-strength either with or without fermions also modifies the magnetic field-amplitude outside the superconductor.  These variations in the external magnetic field thus prevent a direct comparison between the magnetic field transmitted through the superconducting barrier for different wall strengths.  A more suitable measure is the ratio between the magnetic field inside and outside the box $R=B_{\rm out}/B_{\rm in}$, with ``out" defined as halfway between boundary and wall and ``in" is the middle of the box. In practice, we vary the external current to produce a series of ``in"/``out" pairs for different external magnetic fields, and then determine the ratio between the two with a linear fit. 


%

We proceed to calculate this ratio with and without fermions for the range of wall-strengths $\eta$. Given a constant width of the wall $d=1.5$, the fermion contribution should be independent of $\eta$ as it does not couple to the Higgs field. On the other hand, the fermion current clearly notices the change in the gauge field behaviour as it enters the super-conductor (Fig.~\ref{fig:fermcurr}).  

\begin{figure}[H]
  \centering
    \includegraphics[width = 0.8\textwidth]{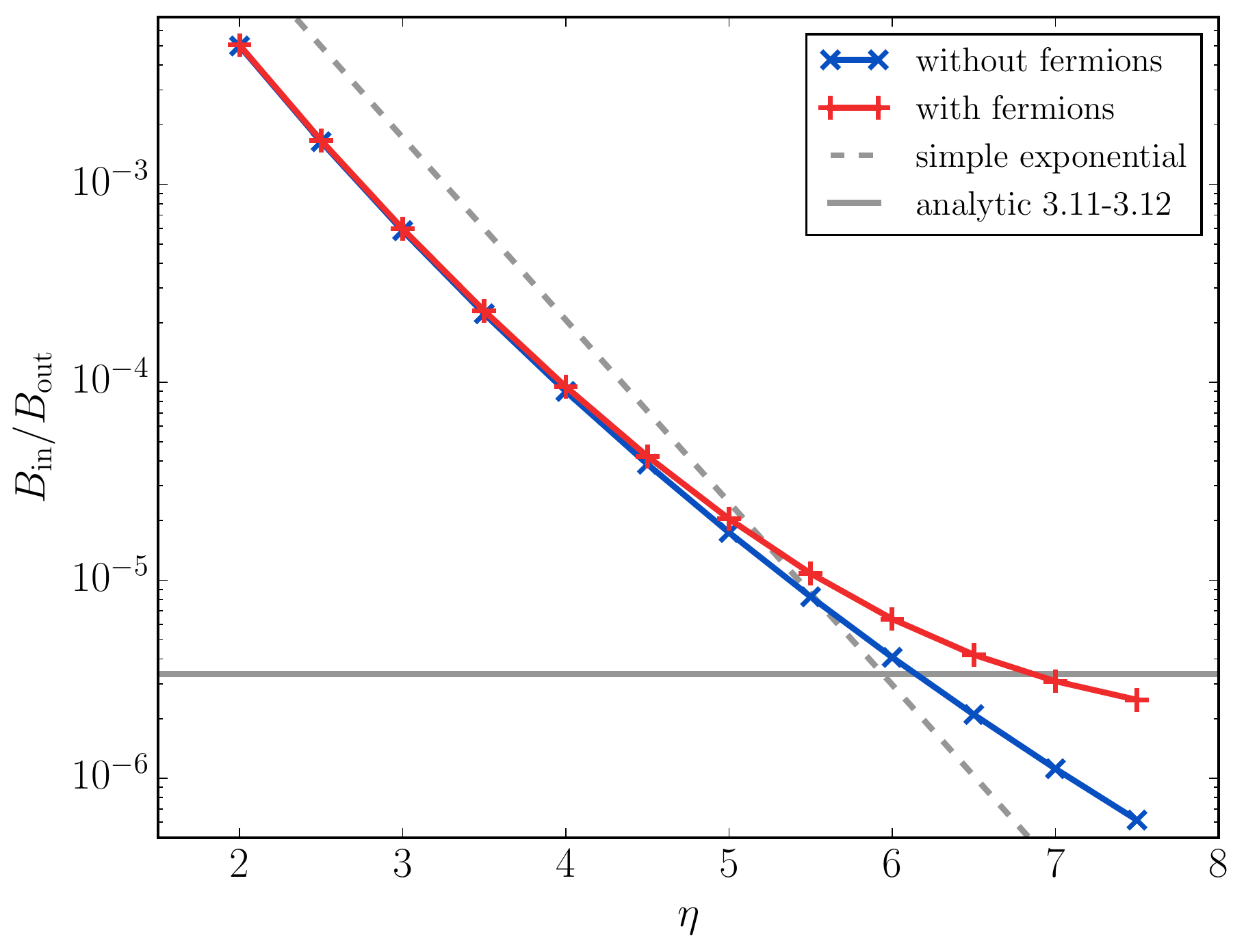}
  \caption{The ratio of magnetic fields inside and outside the box, when varying the wall strength $\eta$ with (red) and without (blue) fermions. Also shown are the analytic estimates (grey solid and grey dashed). Note the log-scale.}
  \label{fig:exp_decay}
\end{figure}

Fig.~\ref{fig:exp_decay} shows the ratio $B_{\rm in}/B_{\rm out}$ as a function of $\eta$, and is our main result.

We see that for small $\eta$, the fermion (red) and no-fermion (blue) results are indistinguishable. As expected, they are not strictly exponential, but the slope is comparable to the simple estimate Eq.~(\ref{eq:exp_est}) (shown in grey, dashed). As we reach $\eta\simeq 5$, the two curves diverge, and the fermion contribution starts to dominate, settling at a constant value of $\simeq 2\times 10^{-6}$ i.e. the analytic estimate for $q=0.3$ and $md=1.5$, of $3.4\times 10^{-6}$, within a factor of two. We consider this a reasonably successful confirmation of that result.


\section{Conclusion}
\label{sec:Conc}
Although not tunnelling in the conventional sense, tunnelling of the 3rd kind is an interesting application of relativistic quantum field effects in a laboratory experiment. We have implemented an idealized laboratory-system of two superconducting slabs shielding a central box from classical electromagnetic fields. Recently developed methods in real-time quantum field theory could then be adapted and refined to successfully simulate the creation of magnetic fields inside the box through this novel tunnelling phenomenon. 

Considering the numerical effort involved and the complexity of detail in the lattice implementation, we are pleasantly surprised that we were able to identify the signal at the level of $10^{-5, -6}$, in accordance with the perturbative prediction within a factor of two. We have relegated many of these details to the appendices, but stress that they are essential for a stable and sufficiently precise computation.

Now that the laboratory is calibrated, it would be interesting to investigate other instances of tunnelling of the 3rd kind, including radiating photons from the sources directly onto the barriers. A specific frequency dependence is expected (see also \cite{third}). Then there is no semi-static magnetic field outside the barriers, and the signal is likely even smaller. Early attempts have been unsuccessful, but are still in progress. Other setups under consideration include different configurations and number of superconducting plates, which will however require larger physical size of the box, with the corresponding increase in numerical effort.

\vspace{0.2cm}

\noindent
{\bf Acknowledgments:}  AT and ZGM are supported by a  UiS-ToppForsk grant from the University of Stavanger. PS and PT acknowledge support by STFC grant ST/L000393/1. The numerical work was performed on 
on the Abel Cluster, owned by the University of Oslo and the Norwegian metacenter for High Performance Computing (NOTUR), and operated by the Department for Research Computing at USIT, the University of Oslo IT-department.

\appendix

\section{Lattice implementation}
\label{sec:LattImp}
In the following, we will provide the details of the lattice action and equations of motion, including how to treat Neumann boundary conditions for our application on a finite lattice in $z$. 

\subsection{Equations of motion}
\label{sec:AppEOM}

The lattice action is discretized as
\begin{align}
S\LATbx = \frac{\DeltaV}{c}\Bigg[&\sum_i \frac{\hbar^2}{2q\mu_0(\Deltax{i})^2(\Deltax{0})^2}\left(2-\Umu[i0]\LATbx-\Umu[0i]\LATbx\right)\nonumber\\
-&\sum_{ij}\frac{\hbar^2}{4q^2\mu_0(\Deltax{i})^2(\Deltax{j})^2}\left(2-\Umu[ij]\LATbx-\Umu[ji]\LATbx\right)\nonumber\\
-&\sum_{\mu}\frac{1}{2\Deltax{\mu}}\bar{\ferm}\LATbx\gamma^{\mu}\left(\Umu\LATbx\ferm\LATbxp{\mu} - \Umu\LATbxm{\mu}\ferm\LATbxm{\mu}\right)\nonumber\\
+&\sum_{i}r_w\frac{1}{2\Deltax{i}}\bar{\ferm}\LATbx\left(\Umu\LATbx\ferm\LATbxp{i} -2\ferm\LATbx + \Umu\LATbxm{i}\ferm\LATbxm{i}\right)\nonumber\\
+&\frac{mc}{\hbar}\bar{\ferm}\LATbx\bar{\ferm}\LATbx\nonumber\\
+&\frac{\hmod^2\LATbx}{2(\Deltax{0})^2}\left(2-\Xphase^\dagger\LATbx\Umu[0]^{e/q}\LATbx\Xphase\LATbxp{0} - \Xphase^\dagger\LATbxp{0}\Umu[0]^{(e/q)\dagger}\LATbx\Xphase\LATbx\right)\nonumber\\
-&\sum_{i}\frac{\hmod^2\LATbx}{2(\Deltax{i})^2}\left(2-\Xphase^\dagger\LATbx\Umu[i]^{e/q}\LATbx\Xphase\LATbxp{i} - \Xphase^\dagger\LATbxp{i}\Umu[i]^{(e/q)\dagger}\LATbx\Xphase\LATbx\right)\Bigg],
\end{align}
in terms of the lattice link $U_\mu$, plaquette $U_{\mu\nu}$ and Higgs field $\Theta$ variables, defined through
\begin{align}
	\Umu\LATbx&\equiv\exp\left(-i\frac{q}{\hbar}\Deltax{\mu} \gauge_{\mu}\LATbx\right) \mbox{,}\\
	\Umu[\mu\nu]\LATbx&\equiv\Umu\LATbx\Umu[\nu]\LATbxp{\mu}\Umu^\dagger\LATbxp{\mu}\Umu[\nu]\LATbx \mbox{,}\\
	\Xphase\LATbx&\equiv\exp\left(i\hphase\LATbx\right)\mbox{.}
\end{align}
Eventually, $\hbar$ and $c$ will be put to unity, but these are included for completeness. $X$ is a lattice point $(x,y,z,t)$, $\Delta x_i$ are the spatial lattice spacings, which may be different although in the main simulations, they are the same,  $\Delta x_1=\Delta x_3=\Delta x_3=0.3$. $\Delta V$ is their product and $\Delta x_0$ is the time-step, which is taken much smaller than the spatial spacings, $0.0002$. $r_w$ is the Wilson coefficient, which will be set to $1/2$, making the fermion lattice doublers massive and drop out of the spectrum. $\pm i$, $\pm j$, $\pm 0$ in the field arguments denotes evaluation at a lattice point shifted in the corresponding spatial or time direction.

Variation with respect to the lattice fields yields the $3+1$D equations of motion, Gauss Law:
\begin{align}
	\sum_i\frac{1}{\Deltax{i}}\frac{1}{\mu_0c}\left(\elec{i}\LATbx-\elec{i}\LATbxm{i}\right) + &\frac{q}{\hbar}\textrm{Im}\{\qaverage[{T\bar{\ferm}\LATbx\gamma^0\ferm\LATbxp{0}}]\} \nonumber\\
	+ &\frac{1}{\Deltax{0}}\frac{e}{\hbar}\hmod^2\LATbx \textrm{Im}\{\Xphase^\dagger\LATbx\Xphase\LATbxp{0}\} =0;
\end{align}
gauge field equation of motion (omitting for now the external current and the damping term):
\begin{multline}
	\label{eq:LAT_EOM_elec_full3d_class}
	\frac{1}{\Deltax{0}}\frac{1}{\mu_0c}\left(\elec{i}\LATbx-\elec{i}\LATbxm{0}\right) \sum_{j}\frac{1}{\Deltax{i}(\Deltax{j})^2}\frac{\hbar}{q\mu_0}\textrm{Im}\left\{\Umu[ij]\LATbx-\Umu[ji]\LATbxm{j}\right\} \\
	-\frac{q}{\hbar}\textrm{Im}\left\{\qaverage[{T\bar{\ferm}\LATbx\gamma^i\Umu[i]\LATbx\ferm\LATbxp{i}}]\right\} + \frac{q}{\hbar}r_w\textrm{Im}\left\{\qaverage[{T\bar{\ferm}\LATbx\Umu[i]\LATbx\ferm\LATbxp{i}}]\right\} \\
	+\frac{1}{\Deltax{i}}\frac{e}{\hbar}\hmod^2\LATbx \textrm{Im}\left\{\Xphase^\dagger\LATbx\Umu[i]^{e/q}\Xphase\LATbxp{i}\right\} = 0;
\end{multline}
fermion field (operator) equation of motion:
\begin{align}
	\frac{1}{2\Deltax{0}}\gamma^{0}\big(&\ferm\LATbxp{0} - \ferm\LATbxm{0}\big) \nonumber\\
	+ \sum_{i}&\frac{1}{2\Deltax{i}}\gamma^{i}\left(\Umu[i]\LATbx\ferm\LATbxp{i} - \Umu[i]^\dagger\LATbxm{i}\ferm\LATbxm{i}\right) + \frac{mc}{\hbar}\ferm\LATbx \nonumber\\ -&\sum_{i}\frac{1}{\Deltax{i}}\frac{r_w}{2}\left(\Umu[i]\LATbx\ferm\LATbxp{i} -2\ferm\LATbx + \Umu^\dagger\LATbxm{i}\ferm\LATbxm{i}\right) = 0;
\end{align}
and Higgs/scalar field equation of motion:
\begin{align}
	\frac{1}{\Deltax{0}}\hmod^2\LATbx\left(\Pi\LATbx - \Pi\LATbxm{0}\right) 
	-\sum_i\frac{1}{(\Deltax{i})^2}\Big(\hmod^2\LATbx \textrm{Im}\left\{\Xphase^\dagger\LATbx\Umu[i]^{e/q}\LATbx\Xphase\LATbxp{i}\right\}& \nonumber\\  
	-\hmod^2\LATbxm{i} \textrm{Im}\left\{\Xphase^\dagger\LATbxm{i}\Umu[i]^{e/q}\LATbxm{i}\Xphase\LATbx\right\}&\Big) = 0.
\end{align}
We have defined the field momenta, for the Higgs field
\begin{equation}
	\Pi\LATbx \equiv \frac{1}{2i\Deltax{0}}\left(\Xphase^\dagger\LATbx\Xphase\LATbxp{0} - \Xphase^\dagger\LATbxp{0}\Xphase\LATbx\right),
\end{equation}
and the electric field
\begin{equation}
\elec{i}\LATbx=\frac{-i}{\Deltax{0}\Deltax{i}}\frac{\hbar c}{2q}\left(\Umu[0i]\LATbx - \Umu[i0]\LATbx\right)\mbox{.}
\end{equation}

After applying the symmetry considerations described in section \ref{sec:Setup}, these reduce to the equation for the $A_1$ gauge field ($T,X_3$ are now time and $z$ coordinate on the lattice, respectively),
\begin{align}
	\label{eq:LAT_EOM_elec_ansatz_quantum}
	&\frac{1}{\Deltax{0}}\frac{1}{\mu_0c}\left(\elec{1}\LATbtz-\elec{1}\LATbtmz\right)\nonumber\\  
	+&\frac{1}{\Deltax{1}(\Deltax{3})^2}\frac{\hbar}{q\mu_0}\textrm{Im}\left\{\Umu[13]\LATbtz-\Umu[31]\LATbtzm{3}\right\}\nonumber\\
	-&\frac{q}{\hbar}\textrm{Im}\left\{\qaverage[{T\qfermbar\LATbx\gamma^1\Umu[1]\LATbtz\qferm\LATbxp{1}}]\right\}
	+ \frac{q}{\hbar}r_w\textrm{Im}\left\{\qaverage[{T\qfermbar\LATbx\Umu[1]\LATbtz\qferm\LATbxp{1}}]\right\} \nonumber\\
	+&\frac{1}{\Deltax{1}}\frac{e}{\hbar}\hmod^2\LATbtz \textrm{Im}\left\{\Umu[1]^{e/q}\LATbtz\right\} = 0\mbox{,}
\end{align}
with
\begin{equation}
\elec{1}\LATbtz=\frac{-i}{\Deltax{0}\Deltax{i}}\frac{\hbar c}{2q}\left(\Umu[0i]\LATbtz - \Umu[i0]\LATbtz\right)\mbox{.}
\end{equation}
The lattice fermions obey 
	\begin{align}
	&\frac{1}{\Deltax{0}}\frac{1}{2}\gamma^0\Big(\fmodez[\LATkas]{A}\LATbtpz - \fmodez[\LATkas]{A}\LATbtmz\Big) \nonumber\\
	+ &\frac{1}{\Deltax{1}}\gamma^1i\sin\left(\Deltax{1}\left[\LATwveci{1} - \frac{q}{\hbar}\Amu[1]\LATbtz\right]\right)\fmodez[\LATkas]{A}\LATbtz \nonumber\\
	+ & \frac{1}{\Deltax{2}}\gamma^2i\sin\left(\Deltax{2}\LATwveci{2}\right)\fmodez[\LATkas]{A}\LATbtz \nonumber\\
	+ &\frac{1}{\Deltax{3}}\frac{1}{2}\gamma^3\left(\fmodez[\LATkas]{A}\LATbtzp{3} - \fmodez[\LATkas]{A}\LATbtzm{3}\right) 
	+ \frac{mc}{\hbar}\fmodez[\LATkas]{A}\LATbtz \nonumber\\
	+ &\frac{1}{\Deltax{1}}r_w\left(\cos\left(\Deltax{1}\left[\LATwveci{1} - \frac{q}{\hbar}\Amu[1]\LATbtz\right]\right)-1\right)\fmodez[\LATkas]{A}\LATbtz \nonumber\\
	+ &\frac{1}{\Deltax{2}}r_w\left(\cos\left(\Deltax{2}\LATwveci{2}\right)-1\right)\fmodez[\LATkas]{A}\LATbtz \nonumber\\
	+ &\frac{1}{\Deltax{3}}\frac{r_w}{2}\Big(\fmodez[\LATkas]{A}\LATbtzp{3} - 2\fmodez[\LATkas]{A}\LATbtz 
	+ \fmodez[\LATkas]{A}\LATbtzm{3}\Big) = 0.
\end{align}
In addition, we have the now redundant constraint-equations
\begin{align}
	%
	- \frac{q}{\hbar}\textrm{Im}\left\{\qaverage[{T\qfermbar\LATbx\gamma^0\qferm\LATbxp{0}}]\right\} + j_0\LATbx&=0\mbox{,}\\
	%
	\frac{q}{\hbar}\textrm{Im}\left\{\qaverage[{T\qfermbar\LATbx\gamma^2\qferm\LATbxp{2}}]\right\} - \frac{q}{\hbar}r_w\textrm{Im}\left\{\qaverage[{T\qfermbar\LATbx\qferm\LATbxp{2}}]\right\}&=0\mbox{,}\\
	%
	\frac{q}{\hbar}\textrm{Im}\left\{\qaverage[{T\qfermbar\LATbx\gamma^3\qferm\LATbxp{3}}]\right\} - \frac{q}{\hbar}r_w\textrm{Im}\left\{\qaverage[{T\qfermbar\LATbx\qferm\LATbxp{3}}]\right\} &=0.
\end{align}

The fermion correlators are
\begin{multline}
	\qaverage[T\qfermbar\LATbx\opgeneric\qferm\LATbxp{0}] = \frac{\hbar c}{2}\sum_{K_1,K_2,\Lambda} \frac{\Delta K_1}{(2\pi)}\frac{\Delta K_2}{(2\pi)}\frac{\Delta \Lambda}{(2\pi)}\\
	\Big(\fmodebarz[\LATkas]{V}\LATbtz\opgeneric\fmodez[\LATkas]{V}\LATbtpz\\ 
	- \fmodebarz[\LATkas]{U}\LATbtz\opgeneric\fmodez[\LATkas]{U}\LATbtpz\Big),
\end{multline}
\begin{multline}
	\qaverage[{T\qfermbar\LATbx\opgeneric\Umu[1]\LATbtz\qferm\LATbxp{1}}] = 
	\frac{\hbar c}{2}\sum_{K_1,K_2,\Lambda} \frac{\Delta K_1}{(2\pi)}\frac{\Delta K_2}{(2\pi)}\frac{\Delta \Lambda}{(2\pi)}e^{i\Deltax{1}\left(\LATwveci{1} - q\Amu[1]\LATbtz\right)}\\
	\Big(\fmodebarz[\LATkas]{V}\LATbtz\opgeneric\fmodez[\LATkas]{V}\LATbtz\\ 
	- \fmodebarz[\LATkas]{U}\LATbtz\opgeneric\fmodez[\LATkas]{U}\LATbtz\Big),
\end{multline}
\begin{multline}
	\qaverage[T\qfermbar\LATbx\opgeneric\qferm\LATbxp{2}] = \frac{\hbar c}{2}\sum_{K_1,K_2,\Lambda} \frac{\Delta K_1}{(2\pi)}\frac{\Delta K_2}{(2\pi)}\frac{\Delta \Lambda}{(2\pi)}e^{i\Deltax{2}\LATwveci{2}}\\
	\Big(\fmodebarz[\LATkas]{V}\LATbtz\opgeneric\fmodez[\LATkas]{V}\LATbtz\\ 
	- \fmodebarz[\LATkas]{U}\LATbtz\opgeneric\fmodez[\LATkas]{U}\LATbtz\Big),
\end{multline}
\begin{multline}
	\qaverage[T\qfermbar\LATbx\opgeneric\qferm\LATbxp{3}] = \frac{\hbar c}{2}\sum_{K_1,K_2,\Lambda} \frac{\Delta K_1}{(2\pi)}\frac{\Delta K_2}{(2\pi)}\frac{\Delta \Lambda}{(2\pi)}\\
	\Big(\fmodebarz[\LATkas]{V}\LATbtz\opgeneric\fmodez[\LATkas]{V}\LATbtzp{3}\\ 
	- \fmodebarz[\LATkas]{U}\LATbtz\opgeneric\fmodez[\LATkas]{U}\LATbtzp{3}\Big)\mbox{.}
\end{multline}

Both the damping and external current are added by hand to match the dynamics in the continuum limit.  The external current explicitly is
\begin{equation}
j_1\LATbx =J_{\rm max} \Theta(T)\left(\exp\left(-\frac{(\LATxi{3}-z_0^2}{2\sigma^2}\right) -\exp\left(-\frac{(\LATxi{3}+z_L)^2}{2\sigma^2}\right)\right),
\end{equation}
defined in terms of the rise function (\ref{eq:LAT_current_time_variation}).  Examining both the discretized fermion and Higgs currents indicates the currents on the lattice typically involve the field values at adjacent lattice-sites and hence the discretized currents naturally correspond to physical locations midway between the sites.  This (and the constraint on the continuum limit) therefore informs the choice to set $z_0 = -\Deltax{3}/2$ and $z_L = (N_Z-1/2)\Deltax{3}$.

\subsection{Boundary conditions}
\label{sec:AppBound}

The Neumann conditions on the $z$ boundaries take the form
\begin{align}
	\partial_z E_1 = \partial_z A_1 = 0,
\end{align}
These imply that the electric field on the lattice satisfies
\begin{align}
	\left.\elec{1}\bx\right|_{n_3=-1} &= \left.\elec{1}\bx\right|_{n_3=0},\\
	\left.\elec{1}\bx\right|_{n_3=N_z} &= \left.\elec{1}\bx\right|_{n_3=N_z-1},
\end{align}
and the gauge field likewise fulfils 
\begin{align}
	\left.\Amu[1]\bx\right|_{n_3=-1} &= \left.\Amu[1]\bx\right|_{n_3=0}, \\
	\left.\Amu[1]\bx\right|_{n_3=N_z} &= \left.\Amu[1]\bx\right|_{n_3=N_z-1}\mbox{.}
\end{align}
The Neumann conditions for the fermion field mean that
\begin{align}
	\partial_z \rho = \partial_z j_{f, 1} = \partial_z j_{f, 2}  =0
	\quad{\rm and}\quad j_{f, 3}=0,
\end{align}
where $\rho$ is the fermion charge density, and $j_{f, 1}$, $j_{f, 2}$, $j_{f, 3}$ the fermion currents along $x$, $y$ and $z$ directions respectively.  (Therefore, no current flows through the $z$ boundaries.)  The equivalent conditions on the lattice may be satisfied through imposing at the two boundaries
\begin{align}
	\ferm\LATbx\big|_{n_3=-1} &= \BfermN{0}\ferm\LATbx\big|_{n_3=0},\\
	\ferm\LATbx\big|_{n_3=N_z} &= \BfermN{N}\ferm\LATbx\big|_{n_3=N_z-1} \mbox{,}
\end{align}
in terms of the constant, $4\times 4$ matrices $\BfermN{0/N}$.  We hence determine $\BfermN{0/N}$ can only be chosen from $\pm i\gamma^5\gamma^3$.

The definition of the Neumann conditions however involve an inherent orientation through the relative sign of the matrices.  Specifically, the operation
\begin{equation}
	\pm i \left(\hat{N}_1\gamma^5\gamma^1 + \hat{N}_2\gamma^5\gamma^2 + \hat{N}_3\gamma^5\gamma^3\right),
\end{equation}
may define the projection of the matrix $\pm i \gamma^5\gamma^i$ on the  outward-orientated, normal vector-field $\mathbf{\hat{N}}\LATbx$ of the lattice. This in particular on the $z$-boundaries implies,
\begin{align}
	\pm i \left(\hat{N}_1\gamma^5\gamma^1 + \hat{N}_2\gamma^5\gamma^2 + \hat{N}_3\gamma^5\gamma^3\right)\Big|_{n_3=-1} &= \mp i\gamma^5\gamma^3 ,\\
	\pm i \left(\hat{N}_1\gamma^5\gamma^1 + \hat{N}_2\gamma^5\gamma^2 + \hat{N}_3\gamma^5\gamma^3\right)\Big|_{n_3=N_z-1} &= \pm i\gamma^5\gamma^3 \mbox{,}
\end{align}
leading to two possible choices:
\begin{equation}
\begin{split}
\BfermN{0} &= +i\gamma^5\gamma^3,\\
\BfermN{N} &= -i\gamma^5\gamma^3,
\end{split}
\quad\mbox{ or }\quad
\begin{split}
\BfermN{0} &= -i\gamma^5\gamma^3,\\
\BfermN{N} &= +i\gamma^5\gamma^3\mbox{.}
\end{split}
\end{equation}
(With these options also, the fermion zero-momentum modes don't exist on the lattice.)  We choose the former for implementation. This constraint in the continuum limit implies further
\begin{equation}
(\Id{4} \mp \BfermN{0/N})\ferm\bx=0,
\end{equation}
at the upper and lower boundaries respectively.  Splitting the fermion spinor into the two-component up-spinor $\ferm_u$ and down-spinor $\ferm_d$ yields
\begin{align}
\ferm_u\bx \pm \sigma^3\ferm_d\bx &=0 ,\nonumber\\
\ferm_d\bx \pm \sigma^3\ferm_u\bx &=0\mbox{.}
\end{align}
These two equations are the same, and thus underdetermine the fermion components. The boundary condition therefore may support an arbitrary fermion-current in the $x$ and $y$ direction on the boundary, consistent with the ansatz.  

\section{Lattice fermions}
\label{sec:AppFerm}

We discretize the fermion field on the lattice taking care of the lattice doublers through a Wilson term. The standard procedure on how this works may be found elsewhere \cite{Robust}.

\subsection{Boundary conditions and mode consistency}
\label{sec:AppFermBound}

Substituting the Neumann vacuum solutions into the mode expansion and applying the ladder commutators with the identification also that initially $\LATmodelabel=\LATwveci{3}$ (though importantly, the weighting $\Delta \Lambda/(2\pi)$ of the integral remains to be determined subsequently) yields:
\begin{align}
	\Big\{\qferm_{a}\LATbzerox[\LATx], &\qferm^{\dagger}_{b}\LATbzerox[\LATy]\Big\} =
	\sum_{\LATwveci{1},\LATwveci{2},\LATwveci{3}} \frac{\Delta K_1}{(2\pi)}\frac{\Delta K_2}{(2\pi)}\frac{\Delta \Lambda}{(2\pi)} e^{i\LATwveci{1}(\LATxi{1} - \LATyi{1})}e^{\LATwveci{2}(\LATxi{2} - \LATyi{2})}\nonumber\\
	&\Bigg[\left(e^{i\LATwveci{3}(\LATxi{3} - \LATyi{3})} + e^{-i\LATwveci{3}(\LATxi{3} - \LATyi{3})}\right)\sum_s \left(\big(U_{\mathbf{\LATwvec},s}\big)_a \big(U^{\dagger}_{\mathbf{\LATwvec},s}\big)_b + \big(V_{-\mathbf{\LATwvec},s}\big)_a \big(V^{\dagger}_{-\mathbf{\LATwvec},s}\big)_b\right)\nonumber\\
	&+ \sum_{c,d} e^{-i\LATwveci{3}(\LATxi{3} + \LATyi{3})}i\left(\gamma^5\right)_{ac}\left(\gamma^3\right)_{cd}\sum_s\left(\big(U_{\mathbf{\LATwvec},s}\big)_d \big(U^{\dagger}_{\mathbf{\LATwvec},s}\big)_b + \big(V_{-\mathbf{\LATwvec},s}\big)_d \big(V^{\dagger}_{-\mathbf{\LATwvec},s}\big)_b\right)\nonumber\\
	&+ \sum_{c,d}e^{i\LATwveci{3}(\LATxi{3} + \LATyi{3})}i\left(\gamma^5\right)_{dc}\left(\gamma^3\right)_{cb}\sum_s\left(\big(U_{\mathbf{\LATwvec},s}\big)_a \big(U^{\dagger}_{\mathbf{\LATwvec},s}\big)_d + \big(V_{-\mathbf{\LATwvec},s}a\big)_a \big(V^{\dagger}_{-\mathbf{\LATwvec},s}\big)_d\right)\Bigg]\mbox{.}
\end{align}
The positive and negative frequency solutions further satisfy the standard result
\begin{align}
	\sum_s U_{\mathbf{\LATwvec},s} U^{\dagger}_{\mathbf{\LATwvec},s} &= \frac{i}{2\omega_{\mathbf{\LATwvec}}}\left(\left(\frac{mc}{\hbar}\right)\Id{4} - i\sum_{\mu}\gamma^{\mu}\LATwveci{\mu}\right)\gamma^{0},\\
	\sum_s V_{\mathbf{\LATwvec},s} V^{\dagger}_{\mathbf{\LATwvec},s} &=\frac{i}{2\omega_{\mathbf{\LATwvec}}}\left(-\left(\frac{mc}{\hbar}\right)\Id{4} - i\sum_{\mu}\gamma^{\mu}\LATwveci{\mu}\right)\gamma^{0}\mbox{,}
\end{align}
and when inserted in the initial condition becomes
\begin{multline}
    \label{eq:fermcorr}
	\left\{\qferm_{a}\LATbzerox[\LATx], \qferm^{\dagger}_{b}\LATbzerox[\LATy]\right\} = \sum_{K_1,K_2,\LATmodelabel} \frac{\Delta K_1}{(2\pi)}\frac{\Delta K_2}{(2\pi)}\frac{\Delta \Lambda}{(2\pi)} e^{i\LATwveci{1}(\LATxi{1} - \LATyi{1})}e^{\LATwveci{2}(\LATxi{2} - \LATyi{1})},\\
	\Bigg(\cos\left(\LATwveci{3}[\LATxi{3} - \LATyi{3}]\right)\left(\Id{4}\right)_{ab} + \cos\left(\LATwveci{3}[\LATxi{3} + \LATyi{3}]\right)\sum_{c,d}i\left(\gamma^5\right)_{ac}\left(\gamma^3\right)_{cb}  \Bigg)\mbox{.}
\end{multline}
Choosing
\begin{equation}
x_3 = \left(n_3 + \frac{1}{2}\right) \Deltax{3} ,
\end{equation}
where $n_3$ is the lattice-site index ensures that these vacuum modes satisfy the Neumann conditions at the origin. We also choose
\begin{equation}
	x_{1,2} = n_{1,2}\Delta x_{1,2} ,
\end{equation}
for the $x$ and $y$ direction, respectively, with periodic boundary conditions.
For the $z$ coordinate, adding two positions implies
\begin{equation}
\LATxi{3} + \LATyi{3} = (n_3 + m_3 + 1)\Deltax{3} \mbox{.}
\end{equation}
This is greater than zero everywhere and hence the second term in the correlator expansion (\ref{eq:fermcorr})  involves a summation over anti-symmetric cosine values,  entirely cancelling pairwise to zero for even $N_z$. 

The $z$ relation similarly implies 
\begin{equation}
\LATxi{3} - \LATyi{3} = (n_3 - m_3)\Deltax{3} \mbox{,}
\end{equation}
non-zero over the lattice except at $\LATxi{3} = \LATyi{3}$, and hence the summation in the $z$-mode direction when $\LATxi{3} \neq \LATyi{3}$ again entirely cancels for even $N_z$ . At $\LATxi{3} = \LATyi{3}$ though, the cosine term equals unity. And equally the summation over the $x$ and $y$-mode directions vanish everywhere except where $\LATxi{1} = \LATyi{1}$ and  $\LATxi{2} = \LATyi{2}$.  The first term in the correlator expansion therefore yields
\begin{equation}
\left\{\qferm_{a}\LATbzerox[\LATx], \qferm^{\dagger}_{b}\LATbzerox[\LATy]\right\} = N_1N_2N_3\frac{\Delta K_1}{(2\pi)}\frac{\Delta K_2}{(2\pi)}\frac{\Delta \Lambda}{(2\pi)}\delta^3_{\mathbf{\LATx}\mathbf{\LATy}}\delta_{ab}\mbox{.}
\end{equation}
Choosing the undetermined $\Deltamode$ weighting to satisfy   
\begin{equation}
\frac{\Delta \Lambda}{(2\pi)} = \frac{\Delta K_3}{\pi} = \frac{\left.\LATwveci{3}\right|_{\kappa_{3+1}} - \left.\LATwveci{3}\right|_{\kappa_{3}}}{\pi},
\end{equation}
explicitly specifies this constant through the lattice wave-vectors.  Substituting the above distance relations into the initial condition, we find
\begin{align}
K_{1,2} &= \frac{2\pi}{L_{1/2}}\left(\kappa_{1/2} + \frac{1}{2}\right),\\
K_{3} &= \frac{\pi}{L_3}\left(\kappa_3 + \frac{1}{2}\right),
\end{align}
where $\kappa_i$ is the site-index in the $i$-direction of the discrete mode-space.  These therefore imply that the initial correlator satisfies
\begin{equation}
\left\{\qferm_{a}\LATbzerox[\LATx], \qferm^{\dagger}_{b}\LATbzerox[\LATy]\right\} = \delta_{ab}\frac{\delta^3_{\mathbf{\LATx}\mathbf{\LATy}}}{\Deltax{1}\Deltax{2}\Deltax{3}}\mbox{.}
\end{equation}
This precisely reproduces the canonical quantization requirements on the discretized fermi\-onic field.

\section{Renormalization}
\label{sec:AppRenorm}

The fermion current on the lattice prior to any renormalization involves both the physical component and the divergent term dependent on the discrete spacing:
\begin{align}
j_{f}\LATbxdx &\equiv 
\bigg(\frac{q}{\hbar}\textrm{Im}\left\{\qaverage[{\qfermbar\LATbx\gamma^1\Umu[1]\LATbtz\qferm\LATbxp{1}}]\right\} \nonumber\\ 
&\quad \quad \qquad -\frac{q}{\hbar}r_w\textrm{Im}\left\{\qaverage[{\qfermbar\LATbx\Umu[1]\LATbtz\qferm\LATbxp{1}}]\right\}\bigg)\bigg|_{\mathbf{\Delta x}} \nonumber\\
&=j_{\rm physical}\LATbxdx + j_{\rm div}\LATbxdx\mbox{.}
\end{align}
We will assume that the lattice-spacing dependence of the physical component is small. In this case, we may to a good approximation write
\begin{equation}
j_{\rm div}\LATbxdx = j_{f}\LATbxdx - j_{f}\LATbxdx[\mathbf{\Delta x_{\rm ref}}] + j_{\rm div}\LATbxdx[\mathbf{\Delta x_{\rm ref}}]\mbox{.}
\end{equation}
Renormalization is achieved by introducing constant counterterms for the operators in the action.  Since the only dynamical equation under consideration is for the gauge field, we will introduce a field renormalization counterterm multiplying the electromagnetic field tensor.  An identical factor for simplicity may be chosen for both the electric field contribution and the term involving purely the plaquettes $\Umu[ij]$. We will introduce these counterterms at the level of the equations of motion. 

This plaquette term in the electric field dynamics corresponds in the continuum limit precisely to the derivative of the magnetic field. The counterterms thus involve a modification to the electric and magnetic field. 

Any lattice-spacing dependence in the renormalization current accordingly occurs with\-in the renormalization coefficient while the spatial variation of the renormalization component occurs entirely in the discretized electromagnetic fields:
\begin{multline}
j_{\rm div}\LATbxdx = \alpha\LATbdx\bigg(\frac{1}{\Deltax{0}}\frac{1}{\mu_0c}\left(\elec{1}\LATbtz-\elec{1}\LATbtmz\right) \\
+\frac{1}{\Deltax{1}(\Deltax{3})^2}\frac{\hbar}{q\mu_0}\textrm{Im}\left\{\Umu[13]\LATbtz-\Umu[31]\LATbtzm{3}\right\}\bigg) \mbox{.}
\end{multline}

The simple case involving a static state and the magnetic gradient identical for each lattice spacings therefore yields
\begin{equation}
\alpha\LATbdx = \Deltax{1}(\Deltax{3})^2\frac{q\mu_0}{\hbar}\frac{j_{f}\LATbxdx - j_{f}\LATbxdx[\mathbf{\Delta x_{ref}}]}{\textrm{Im}\left\{\Umu[13]\LATbtz-\Umu[31]\LATbtzm{3}\right\}} + \alpha\LATbxdx[\mathbf{\Delta x_{ref}}]\mbox{.}
\end{equation}
We choose $\alpha\LATbxdx[\mathbf{\Delta x_{ref}}] = 0$.  For comparison, the perturbative calculation for our choice parameters yields a value of $\alpha\LATbxdx[\mathbf{\Delta x_{ref}}] = 0.004$, which is a small correction to the effective gauge coupling.

Obtaining the renormalization coefficient, in practice, is accomplished through evolving the fermion fields from the initial vacuum-configuration to the final, static state with the gauge field set manually (and without any Higgs-field).  
Substituting the external current on the lattice into the continuum gauge dynamics determines the magnetic field in the static state satisfies
\begin{equation}
\frac{1}{\mu_0}\partial_3\magni{2}\btz = \frac{1}{\mu_0}\partial^2_3\Amu[1]\btz = |J|\exp\left(-\frac{(z-z_0)^2}{2\sigma^2}\right) -|J|\exp\left(-\frac{(z-z_L)^2}{2\sigma^2}\right)\mbox{.}
\end{equation}
Integrating this expression we find
\begin{align}
\Amu[1]\btz = &\mu_0\sqrt{\frac{\pi}{2}}\sigma|J|\Bigg[(z-z_0)\erf\left(-\frac{(z-z_0)^2}{2\sigma^2}\right) - (z+z_L)\erf\left(-\frac{(z+z_L)^2}{2\sigma^2}\right) \nonumber\\
&+ \sqrt{\frac{2}{\pi}}\sigma\exp\left(-\frac{(z-z_0)^2}{2\sigma^2}\right) -  \sqrt{\frac{2}{\pi}}\sigma\exp\left(-\frac{(z+z_L)^2}{2\sigma^2}\right)\Bigg] + B_Cz + A_C\mbox{,}
\end{align}
where 
\begin{align}
B_C = &-\mu_0\sqrt{\frac{\pi}{8}}\sigma|J|\left[\erf\left(-\frac{(z_M-z_0)^2}{2\sigma^2}\right) - \erf\left(-\frac{(z_M+z_L)^2}{2\sigma^2}\right)\right]\\
A_C = &-\mu_0\sqrt{\frac{\pi}{2}}\sigma|J|\Bigg[(z_M-z_0)\erf\left(-\frac{(z_M-z_0)^2}{2\sigma^2}\right) - (z_M+z_L)\erf\left(-\frac{(z_M+z_L)^2}{2\sigma^2}\right) \nonumber\\
&+ \sqrt{\frac{2}{\pi}}\sigma\exp\left(-\frac{(z_M-z_0)^2}{2\sigma^2}\right) -  \sqrt{\frac{2}{\pi}}\sigma\exp\left(-\frac{(z_M+z_L)^2}{2\sigma^2}\right)\Bigg] - B_Cz_M\\
z_M =& \frac{z_0 + z_L}{2}\mbox{.}
\end{align}
The constants are chosen to ensure the analytic magnetic and gauge fields viewed along the $z$-direction form a symmetrical configuration centred on the $z$-axis.  Converting the coordinates trivially to lattice positions and the rise function (\ref{eq:LAT_current_time_variation}) multiplying this static configuration consequently provides the gauge-field background for the fermion evolution.

\begin{figure}[H]
	\centering
	\includegraphics[width = 0.45\textwidth]{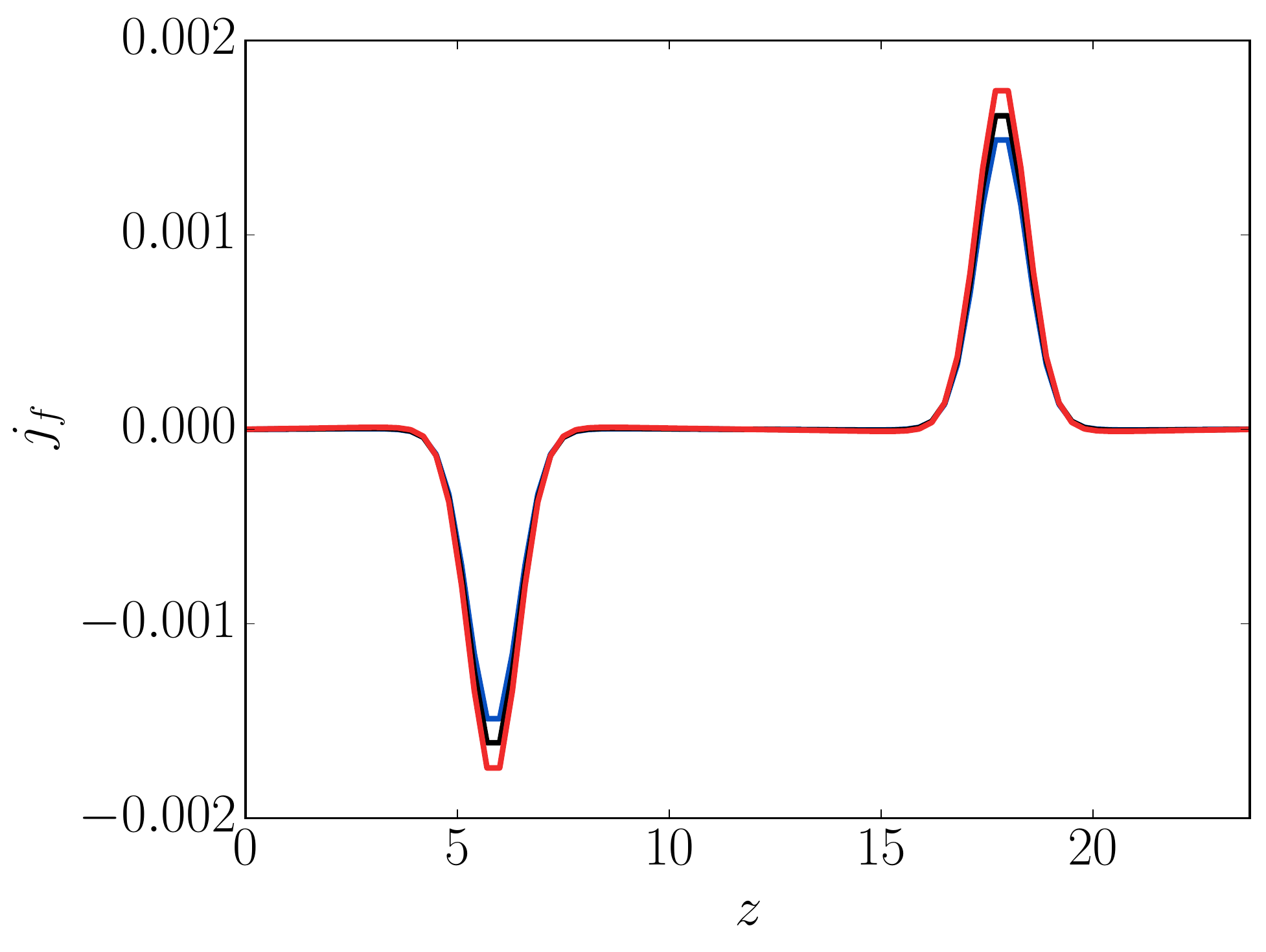}
	\includegraphics[width = 0.45\textwidth]{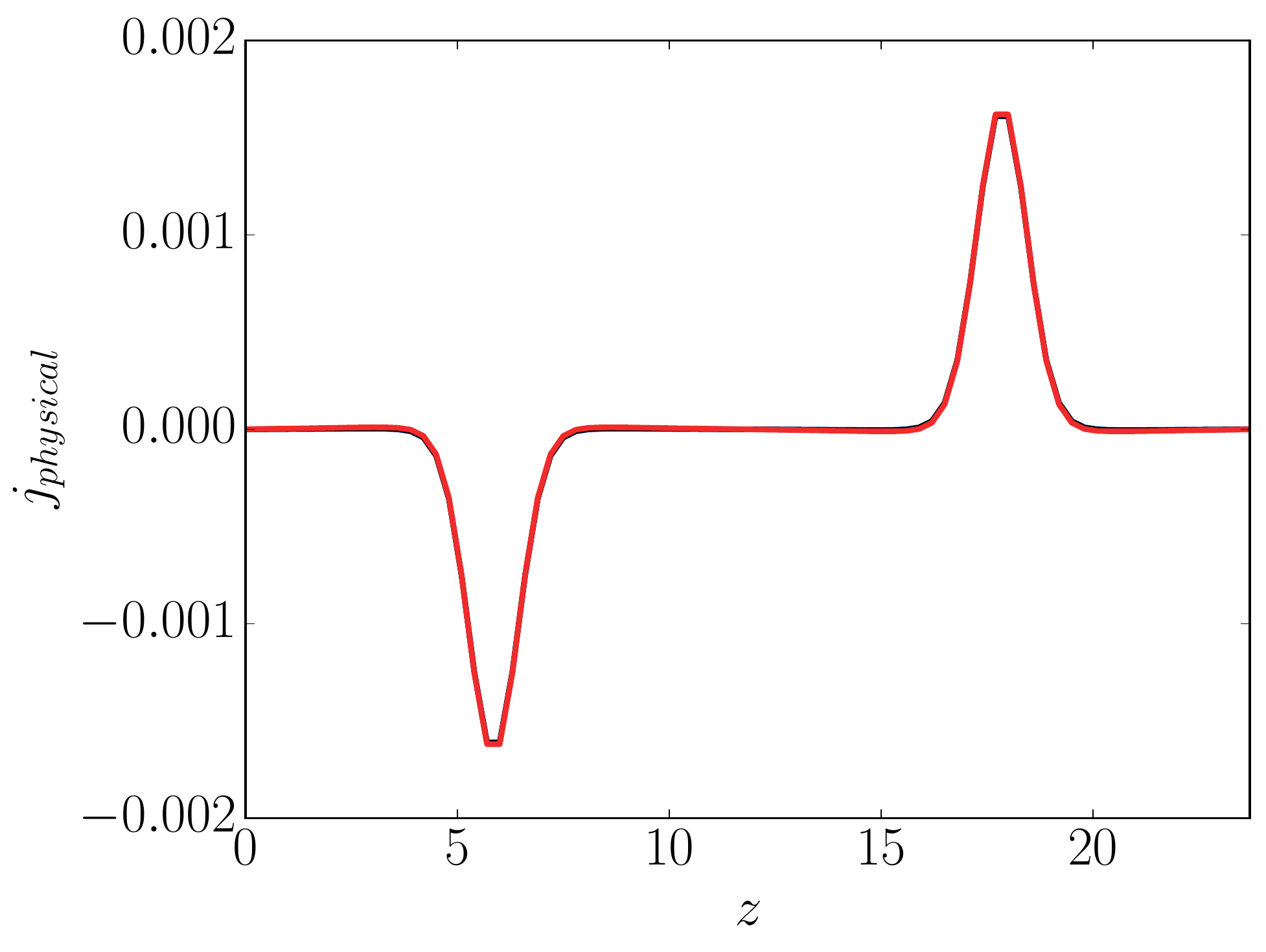}
	\caption{The bare current $j_f$ (left) and the renormalized current (right) for $\Deltax{1}=0.4$ (blue), 0.3 (black), 0.2 (red).  $\alpha = 0$ for $\Deltax{1} = 0.2$ and $\alpha = 0$ for $\Deltax{1} = 0.2$; $\Deltax{2} = 0.3$ is the reference lattice spacing where $\alpha \equiv 0$. $N_x\Deltax{1} = N_y\Deltax{y} = 9.6$, $N_z\Deltax{3} = 9.6$,  $q=0.3$, $e=1$ and periodic boundaries are applied in all directions.}
	\label{fig:renorm}
\end{figure}

Obtaining the renormalization coefficients hence enables $j_{\rm physical} = j_f - j_{\rm div}$ to replace the bare current in the fully dynamical case.  The Fig. \ref{fig:renorm} shows the convergence (right) on implementing the renormalization scheme for $\mathbf{\Delta x_{ref}} = (0.3, 0.3, 0.3)$ with $\Deltax{2} = \Deltax{3} = 0.3$ fixed and $\Deltax{1}$ varying, and we see that the curves for the renormalized current lie on top of each other.  


\end{document}